\documentclass[journal]{IEEEtran}

\usepackage{caption}
\usepackage{amsfonts}
\usepackage{amssymb}
\usepackage{stfloats}
\usepackage{cite}
\usepackage{graphicx}
\usepackage{epstopdf}
\usepackage{psfrag}
\usepackage{subfigure}
\usepackage{amsmath}
\usepackage{array}
\usepackage{stfloats}
\usepackage{multirow}
\usepackage{color}
\usepackage{booktabs}
\usepackage{url}
\usepackage{graphicx}
\usepackage{enumitem}
\usepackage{lipsum} % 调用跨栏长公式宏包

\usepackage{amsmath}
\usepackage{amssymb}
\usepackage{tabularx}
\usepackage{booktabs} 
\usepackage{marvosym} 
\usepackage{url}
\usepackage{hyperref}
\usepackage{makecell}
\usepackage{bm}

\usepackage[linesnumbered,ruled,vlined]{algorithm2e}

\begin{document}

% \title{A Novel Reflector-RBF Approach for IREE Oriented RIS-aided 6G Network}

% \title{Cross-Attention Radial Basis Function Network for IREE Oriented RIS-Assisted 6G Network Design} Capacity Constrained Green Wireless Network Scaling: Multi-BS Collaboration or Multi-RIS Collaboration?

\title{DyTransRM: A Unified Simulation-to-Real Framework for Dynamic Radio Map Prediction}

\author{\IEEEauthorblockN{Simin~Wang\IEEEauthorrefmark{1},
Jiayi~Wei\IEEEauthorrefmark{1},
Jihong~Li\IEEEauthorrefmark{1},
Rui~Cheng\IEEEauthorrefmark{1},
Shunqing~Zhang\IEEEauthorrefmark{1}}

\IEEEauthorblockA{\IEEEauthorrefmark{1}School of Information and Communication Engineering, Shanghai University, Shanghai, 200444, China \\}

\thanks{This work was supported by the National Natural Science Foundation
of China under Grant 62571307, the Science and Technology Commission
Foundation of Shanghai under Grants 24DP1500500 and the
National Key Research and Development Program of China under Grant
2022YFB2902304.}
\thanks{Corresponding Author: {\em Shunqing Zhang}.}

\thanks{Simin Wang, Jiayi Wei, Jihong Li, Rui Cheng and Shunqing Zhang are with Shanghai Institute for Advanced Communication and Data Science, Key Laboratory of Specialty Fiber Optics and Optical Access Networks, Shanghai University, Shanghai, 200444, China (e-mails: \{siminwang, weijiayi, tomlijiong, cheneycheng, shunqing\}@shu.edu.cn).}
}

\markboth{Journal of \LaTeX\ Class Files,~Vol.~14, No.~8, August~2015}%
{Shell \MakeLowercase{\textit{et al.}}: Bare Demo of IEEEtran.cls for IEEE Journals}

\maketitle

\begin{abstract}
Dynamic radio map (RM) prediction enables proactive wireless network control by forecasting future propagation conditions from historical observations. However, collecting temporally continuous and spatially dense real-world RMs is prohibitively expensive, while models trained on large-scale simulated data often suffer from limited transferability to real environments. To address this issue, we propose DyTransRM, a unified simulation-to-real framework for dynamic RM prediction under sparse real-world measurements. DyTransRM first introduces RadioCLIP to learn physics-semantic representations by aligning environmental evolution with the corresponding propagation evolution, thereby improving both prediction accuracy and cross-domain transferability. To further bridge the simulation-to-real gap, SparseUOT employs sparse-mask simulation refinement to mimic realistic measurement patterns and unbalanced optimal transport to selectively align compatible propagation representations across domains. Experiments on RadioMapMotion and real-world datasets demonstrate that RadioCLIP achieves an NMSE of 0.0018 and an RMSE of 0.012 in simulation-domain prediction. After sparse real-domain adaptation, DyTransRM achieves an NMSE of 0.023, RMSE of 0.079, and MAE of 0.064, outperforming representative adaptation baselines without requiring spatially complete real RMs. These results demonstrate the effectiveness of DyTransRM for transferring simulation-learned dynamic propagation knowledge to practical sparse-measurement scenarios.
\end{abstract}


\section{Introduction} \label{sect:intro}

Radio maps (RMs) provide spatially continuous representations of wireless propagation conditions over a target area. They can offer prior channel knowledge for beam management, resource allocation, and other network control tasks without relying on instantaneous pilot-based channel measurements\cite{10430216}. However, the validity of such prior knowledge strongly depends on the stability of the physical environment. In vehicular networks, low-altitude communications, and other highly dynamic scenarios, moving scatterers such as vehicles and unmanned aerial vehicles continuously alter blockage, reflection, and diffraction conditions, causing the radio propagation field to evolve over time\cite{li2023channel}. Meanwhile, next-generation wireless networks are increasingly moving toward environment-aware operation and digital-twin assisted intelligence. This evolution requires not only accurate but also forward-looking knowledge of the radio environment\cite{10577097}. Therefore, reconstructing only the current RM is no longer sufficient for proactive network control. Predicting future spatial propagation states from historical observations has become an important capability for proactive radio environment awareness\cite{11433407}. However, this objective introduces a fundamental contradiction: \emph{Acquiring continuous spatio-temporal propagation information requires extensive channel measurements, whereas only very limited measurements are available in practical wireless networks.}

Existing studies have explored different approaches to predict the spatio-temporal evolution of radio propagation from limited measurements. Early works mainly adopted interpolation-based methods, such as Kriging, Gaussian processes, and Kalman filtering, to reconstruct complete radio maps from sparse observations~\cite{maeng2024kriging,chen2025high,qi2025data}. Although these methods can exploit local spatial correlations, they generally rely on predefined statistical assumptions and simplified noise models, making them insufficient to characterize the complex propagation variations caused by dynamic environments. With the rapid development of deep learning, recent studies have introduced large-scale neural networks, including Transformer-based architectures, ConvLSTM, and BERT, to directly learn nonlinear spatio-temporal propagation patterns from data~\cite{quang20253d,cheng2026radiomapmotion,zhao2025csi}. These approaches significantly improve prediction accuracy by leveraging large-scale training datasets, but their performance heavily depends on the availability of massive labeled propagation data, which is difficult to obtain from real-world measurements. To alleviate this data acquisition bottleneck, recent studies have increasingly adopted model-data-driven approaches that combine physical simulation and data-driven learning, where ray-tracing platforms are employed to generate large-scale propagation datasets for model training~\cite{cheng2026radiomapmotion}. However, the discrepancy between simulated and real propagation environments limits the direct deployment of simulation-trained models in practical scenarios.

Although existing simulation-to-real adaptation methods, including fine-tuning, meta-learning, and adversarial learning~\cite{liao2025visual,zeng2025meta,muruganandham2025smart,lu2026transfer}, have achieved promising results, they mainly rely on dense spatial measurements collected by fixed measurement devices at multiple locations. However, in dynamic scenarios, the additional temporal dimension further increases measurement sparsity, making existing sim-to-real adaptation methods based on dense spatial measurements ineffective.

To address the above challenges, we propose DyTransRM, a unified
simulation-to-real framework for dynamic RM prediction under sparse
real-world measurements. The core idea is to first pretrain the model
on large-scale simulated data and subsequently adapt it to real-world
scenarios using only a limited number of measurements. Unlike
conventional simulation-based pretraining methods that primarily
rely on pixel-level RM fitting, DyTransRM introduces RadioCLIP
to learn transferable propagation semantics by explicitly capturing
the correspondence between environmental evolution and propagation
evolution. Building upon this semantic-aware pretrained model,
SparseUOT further facilitates simulation-to-real adaptation through
sparse-observation training and unbalanced optimal transport (UOT),
enabling effective knowledge transfer under severe measurement
sparsity and cross-domain distribution discrepancies. Together,
these designs enable accurate future RM prediction without
requiring spatially complete real-world RM labels.
The main contributions of this work are summarized as follows.

\begin{itemize}

    \item{\em A Unified Simulation-to-Real Framework for Dynamic RM Prediction.}
    We develop DyTransRM to address the emerging challenge of transferring simulation-learned dynamic propagation knowledge to real-world environments, where continuous RM acquisition is impractical and only sparse observations are available. Unlike existing studies that mainly consider simulation-domain prediction or static RM transfer, DyTransRM jointly tackles dynamic RM prediction, sparse real-world measurements, and simulation-to-real distribution discrepancies, enabling future propagation prediction from large-scale simulation data under practical deployment constraints.

    \item{\em Physics-Semantic Learning of Transferable Propagation Knowledge.}
    To address the challenge of aligning dynamic environmental changes with propagation evolution, we propose RadioCLIP, a physics-semantic learning mechanism based on causal cross-modal contrastive learning. By capturing the correspondence between environmental evolution and propagation evolution, RadioCLIP learns transferable dynamic propagation representations. Experimental results demonstrate that RadioCLIP improves the RMSE performance by 20\% compared with representative dynamic RM prediction methods on large-scale simulated datasets, while maintaining effective transferability to real-world scenarios.

    \item{\em Sparse Observation-Aware Simulation-to-Real Adaptation via UOT Alignment.}
    To address the challenges of severe measurement sparsity and distribution discrepancies between simulated and real propagation environments, we propose SparseUOT, which combines sparse-mask simulation refinement with unbalanced optimal transport to enable adaptation under deployment-like sparse observations. By selectively aligning transferable propagation representations across domains, SparseUOT effectively bridges the simulation-to-real gap. Extensive experiments demonstrate that SparseUOT improves adaptation performance by 24\% under sparse real-world measurements.

    \item{\em Comprehensive Performance Analysis and Deployment Evaluation.}
    We conduct extensive experiments to validate the effectiveness and practicality of DyTransRM from multiple perspectives, including component-wise ablation studies, comparisons between balanced and unbalanced optimal transport, computational complexity analysis, and cross-dataset evaluation. The experiments are conducted on large-scale simulated and real-world datasets, including RadioMapMotion, Vienna, and Wiworld datasets, demonstrating the robustness and practical applicability of the proposed framework across diverse propagation scenarios.
\end{itemize}

The remainder of this paper is structured as follows. Section~\ref{sect:related_work} presents the related work. Section~\ref{sect:system_model} details the system model and formulates the RM prediction problem. Section~\ref{sect:algorithm} presents the DyTransRM framework as well as its constituent components: RadioCLIP and the SparseUOT scheme. Section~\ref{sect:num_res} provides numerical results, and Section~\ref{sect:conc} concludes the paper.

\section{Related Work}
\label{sect:related_work}

\subsection{Radio Map Construction and Dynamic Prediction}

Radio map construction has evolved from measurement-driven spatial interpolation toward data-driven propagation learning. Deep completion models exploit spatial correlations to reconstruct RMs from sparse measurements, while RadioUNet incorporates environmental layouts to learn propagation maps from large-scale simulated data~\cite{teganya2020data,levie2021radiounet}. Public ray-tracing datasets such as RadioMapSeer have facilitated learning-based RM modeling at scale~\cite{yapar2023dataset}. Subsequent studies have explored U-Net, generative, and Transformer-based architectures to improve RM reconstruction and prediction under different propagation environments~\cite{chaves2023deeprem,krijestorac2023agile,wang2024radiodiff,li2025transpathnet}.  More recently, research has extended RM modeling from static reconstruction to temporal prediction. RadioMapMotion introduced a large-scale benchmark for proactive spatio-temporal radio environment prediction~\cite{cheng2026radiomapmotion}, while vision Transformer-based models have also been explored for predicting dynamic 3D RMs~\cite{quang20263d}. These studies demonstrate the potential of deep spatio-temporal modeling, but existing dynamic approaches primarily learn propagation evolution directly from RM sequences.

\subsection{Representation Learning for Wireless Propagation}

Recent wireless learning has increasingly shifted from task-specific prediction toward transferable representation learning. Masked reconstruction has been employed to learn generalizable channel representations, as exemplified by CSI-MAE~\cite{jiang2026csi}, while JEPA-CFM learns high-level latent channel structures through joint-embedding predictive learning~\cite{gao2026jepa}. More broadly, contrastive multimodal learning, represented by CLIP~\cite{radford2021learning}, has demonstrated that aligning paired modalities can produce transferable semantic representations. These advances motivate learning propagation representations beyond raw-value reconstruction. Different from existing approaches that primarily model channel observations themselves, RadioCLIP exploits the physical correspondence between environmental evolution and RM evolution to learn environment-conditioned propagation semantics.

\subsection{Simulation-to-Real Transfer and Domain Alignment}

Simulation provides scalable propagation data but inevitably introduces a distribution gap with real measurements. Transfer learning has therefore been investigated for RM prediction, where knowledge learned from one propagation configuration is adapted using limited target-domain observations~\cite{parera2020transfer}. More generally, domain adaptation methods seek domain-invariant representations through distribution alignment, among which optimal transport (OT) provides a principled mechanism for matching source and target distributions~\cite{courty2017optimal}. DeepJDOT further integrates OT with deep representation learning~\cite{damodaran2018deepjdot}, while unbalanced OT relaxes strict mass-preservation constraints and has shown improved robustness for domain adaptation~\cite{fatras2021unbalanced}. Inspired by these developments, SparseUOT employs unbalanced transport to selectively align compatible simulated and real propagation representations under sparse real-world observations.

\section{System Model and Problem Formulation}
\label{sect:system_model}

This section introduces the system model for dynamic radio map prediction, formulates the multi-step prediction problem under sparse observations, and highlights the key challenges associated with real-world deployment.

\subsection{System Model}

We consider a target area discretized into an $N\times N$ grid. 
At discrete time slot $t\in\{1,\ldots,T\}$, the corresponding radio map 
(RM) is represented as
\begin{equation}
P_t = [p_{i,j,t}]_{i,j=1}^{N} \in \mathbb{R}^{N\times N},
\end{equation}
where $p_{i,j,t}$ denotes the path loss at the $(i,j)$-th grid cell 
at time slot $t$. The environmental context at time slot $t$ is characterized by an 
occupancy map
\begin{equation}
H_t = [h_{i,j,t}]_{i,j=1}^{N} \in \{0,1\}^{N\times N},
\end{equation}
where $h_{i,j,t}=1$ indicates that the $(i,j)$-th grid cell is occupied 
by an environmental object, such as a building or a vehicle, whereas 
$h_{i,j,t}=0$ indicates an unoccupied cell. The BS location is represented by a binary map
\begin{equation}
R_t = [r_{i,j,t}]_{i,j=1}^{N} \in \{0,1\}^{N\times N},
\end{equation}
where $r_{i,j,t}=1$ if the $(i,j)$-th grid cell corresponds to the BS 
location and $r_{i,j,t}=0$ otherwise. The complete environmental representation is denoted by
\begin{equation}
C_t = \left[H_t, R_t\right]  \in \mathbb{R}^{N\times N\times 2}.
\end{equation}

\subsection{Problem Formulation}

We divide the temporal sequence into a historical observation interval
and a future prediction interval. Specifically, the sparse RM
observations from time slot $1$ to $\hat{T}$ are available as historical
propagation information, while the RMs from time slot $\hat{T}+1$ to
$T$ are to be predicted, where $1 \leq \hat{T} < T$.

For notational convenience, we define the environmental sequence and the historical RM observations as
\begin{equation}
\mathcal{C}_{1:\hat{T}}
\triangleq
\{C_t\}_{t=1}^{\hat{T}},
\end{equation}
\begin{equation}
\mathcal{P}_{1:\hat{T}}
\triangleq
\{P_t\}_{t=1}^{\hat{T}},
\end{equation}
respectively.

Given $\mathcal{C}_{1:\hat{T}}$ and
$\mathcal{P}_{1:\hat{T}}$, the objective is to estimate the future RM sequence $\widehat{\mathcal{P}}_{\hat{T}+1:T}$, and the parameterized predictor $\mu_{\theta}$ is formulated as
\begin{equation}
\widehat{\mathcal{P}}_{\hat{T}+1:T}
=
\mu_{\theta}
\left(
\mathcal{C}_{1:\hat{T}},
\mathcal{P}_{1:\hat{T}}
\right),
\label{eq:dynamic_rm_prediction}
\end{equation}
where $\theta$ denotes the learnable model parameters.

The prediction problem can then be formulated as
\begin{equation}
\min_{\theta}
\frac{1}{T-\hat{T}}
\sum_{t=\hat{T}+1}^{T}
\frac{1}{|\Omega_t|}
\sum_{(i,j)\in\Omega_t}
\ell
\left(
\widehat{p}_{i,j,t},
p_{i,j,t}
\right),
\label{eq:problem_formulation}
\end{equation}
where $\ell(\cdot,\cdot)$ denotes the point-wise prediction loss and $|\Omega_t|$ represents the number of available measurements at time
slot $t$. Realizing such a predictor in practical wireless networks involves two
fundamental challenges.
First, Although existing studies have explored methods for dynamically predicting RM on large-scale simulation datasets, they typically focus on pixel-to-pixel evolution rather than the learning of wireless propagation knowledge, resulting in limited prediction performance. Second, real-world dynamic RM measurements are extremely sparse. Unlike static RMs, measurements collected at different time slots cannot be freely aggregated because the propagation field evolves with the environment. As a result, only a few spatial observations are available at each timeslot, making direct data-driven training difficult. Third, a substantial simulation-to-real domain gap remains. Although ray tracing can generate large-scale dynamic RM data, simplified map geometry derived from OpenStreetMap lead to systematic discrepancies between simulated and measured RMs. Therefore, models trained purely in simulation may generalize poorly to real environments.

\begin{figure*}[t]
\centering
\includegraphics[width=1\textwidth]{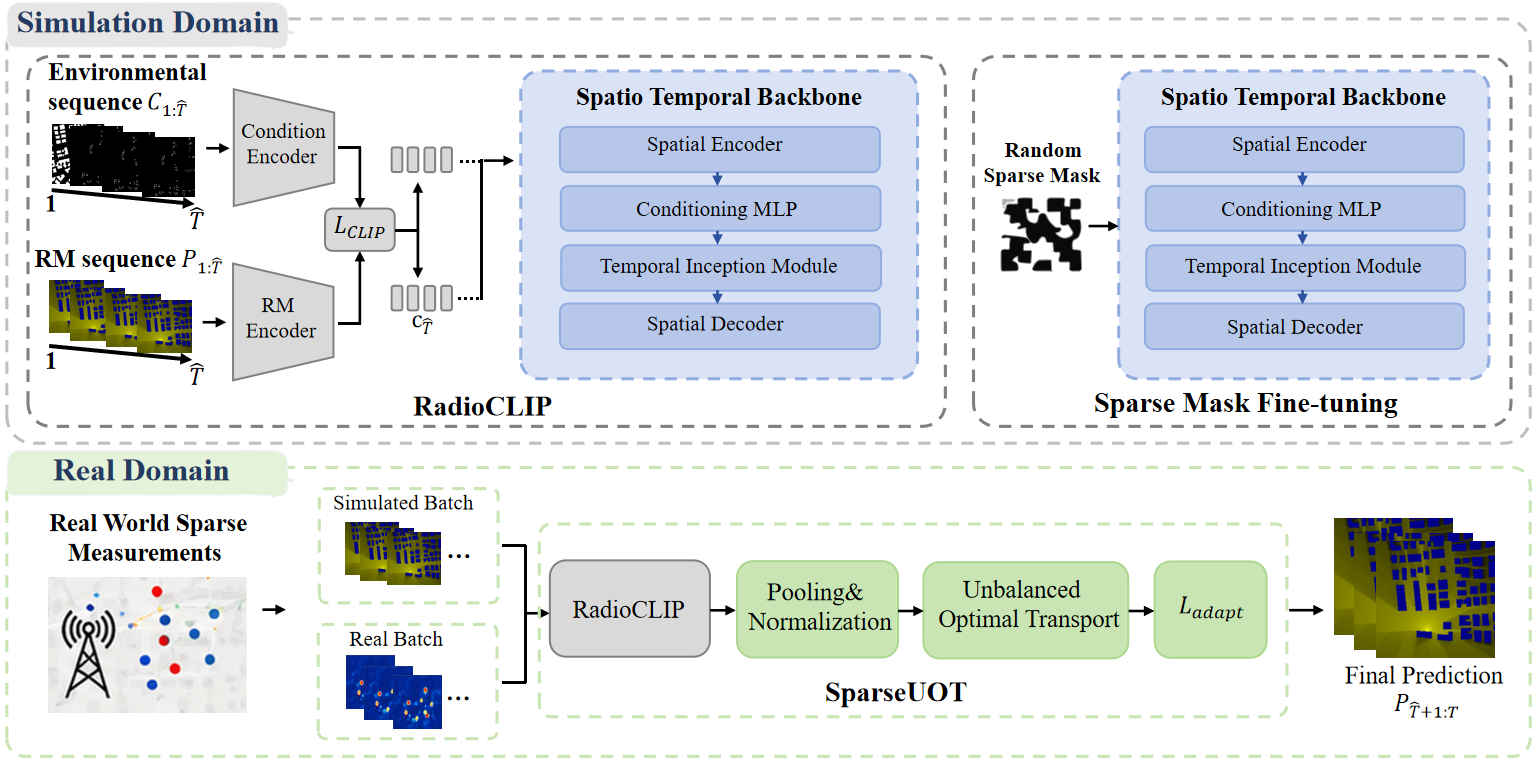}
\caption{Overview of the proposed DyTransRM framework. 
RadioCLIP aligns environmental and RM sequences to learn physics-semantic conditions for spatio-temporal prediction, while SparseUOT aligns simulated and real propagation representations for effective simulation-to-real adaptation.}
\label{fig:overall_framework}
\end{figure*}

\section{Proposed DYTRANSRM Framework}
\label{sect:algorithm}

In this section, we present the proposed DyTransRM framework for dynamic RM prediction under sparse real-world measurements. We first introduce RadioCLIP, using causal cross-modal contrastive learning to align environmental evolution with propagation evolution, thereby extracting transferable physics-semantic representations for dynamic RM prediction. Moreover, to address the severe measurement sparsity and simulation-to-real distribution discrepancy in practical deployment, SparseUOT first adopts sparse-mask simulation refinement to mimic realistic observation patterns during training, and then employs unbalanced optimal transport to selectively align compatible propagation representations between simulation and real domains. Through the joint design of physics-semantic learning and sparse observation-aware domain adaptation, DyTransRM enables effective transfer of simulation-learned dynamic propagation knowledge to real-world scenarios with limited measurements.

\subsection{Framework Overview}
\label{subsec:framework_overview}

DyTransRM follows a simulation-pretraining and real-domain adaptation paradigm. 
The basic idea is to first exploit large-scale simulated RM sequences to learn an accurate and transferable dynamic RM predictor, and then adapt the simulation-learned knowledge to practical environments using only sparse real measurements. 
To this end, DyTransRM incorporates two key components. 
RadioCLIP learns environment-conditioned propagation semantics by explicitly associating environmental evolution with the corresponding RM evolution, and uses the learned physics-semantic representation to guide future RM prediction. 
SparseUOT further exposes the predictor to deployment-like sparse observations during simulation training and selectively aligns compatible propagation representations between the simulation and real domains through unbalanced optimal transport. 
The overall architecture of DyTransRM is illustrated in Fig.~\ref{fig:overall_framework}.
\subsection{RadioCLIP: Physics-Semantic Learning for Dynamic RM Prediction}
\label{subsec:radioclip}

Dynamic RMs are determined by both the temporal evolution of the propagation
field and the corresponding changes in the surrounding environment.
Modeling RM sequences solely through pixel-level temporal correlations may
therefore lead the predictor to capture dataset-specific patterns rather
than stable environment--propagation relationships. To address this issue,
we develop RadioCLIP, which consists of two stages. First, a CLIP-based
cross-modal learning module is pretrained to extract propagation-relevant
semantics from historical environmental evolution. The resulting
physics-semantic representation is then directly incorporated into a
spatio-temporal RM prediction network to infer future RMs.

%%因果transformer再过渡得丝滑一点。

\textit{1) CLIP-Based Cross-Modal Physics-Semantic Pretraining:} For each simulated training sample, the historical environmental sequence
and its corresponding RM sequence are defined as
\begin{equation}
\mathcal{C}_{1:\hat{T}}
\triangleq
\{C_t\}_{t=1}^{\hat{T}},
\qquad
\mathcal{P}_{1:\hat{T}}
\triangleq
\{P_t\}_{t=1}^{\hat{T}},
\label{eq:clip_sequences}
\end{equation}
where only observations up to time slot $\hat{T}$ are involved in the
cross-modal learning process.

To learn their temporal correspondence, we employ two
causal Transformer encoders for the environmental and propagation
branches, respectively. Each frame is first partitioned into
non-overlapping spatial patches and projected into patch tokens, followed
by spatial and temporal positional embeddings. Causal self-attention is
then applied along the temporal dimension, such that the representation at
each time slot is constructed only from the current and preceding
environmental or propagation states\cite{lan2026stream3r}. Accordingly,
\begin{equation}
\mathbf{H}_{t}^{c}
=
f_{\psi}
\left(
\mathcal{C}_{1:t}
\right),
\qquad
\mathbf{H}_{t}^{p}
=
g_{\phi}
\left(
\mathcal{P}_{1:t}
\right),
\quad
1\leq t\leq\hat{T},
\label{eq:causal_cross_modal_encoding}
\end{equation}
where $f_{\psi}(\cdot)$ and $g_{\phi}(\cdot)$ denote the environmental
and propagation encoders with parameters $\psi$ and $\phi$, respectively.
This causal encoding preserves the temporal direction of the two
modalities and provides temporally consistent representations for
subsequent cross-modal alignment.

For contrastive learning, the representations at the end of the historical
interval are spatially aggregated:
\begin{equation}
\bar{\mathbf{h}}^{c}
=
\operatorname{Pool}
\left(
\mathbf{H}_{\hat{T}}^{c}
\right),
\qquad
\bar{\mathbf{h}}^{p}
=
\operatorname{Pool}
\left(
\mathbf{H}_{\hat{T}}^{p}
\right),
\label{eq:clip_pooling}
\end{equation}
where $\operatorname{Pool}(\cdot)$ denotes spatial token pooling. The
resulting representations are projected into a shared latent space:
\begin{equation}
\mathbf{z}^{c}
=
\frac{
\mathbf{W}_{c}\bar{\mathbf{h}}^{c}
}{
\left\|
\mathbf{W}_{c}\bar{\mathbf{h}}^{c}
\right\|_{2}
},
\qquad
\mathbf{z}^{p}
=
\frac{
\mathbf{W}_{p}\bar{\mathbf{h}}^{p}
}{
\left\|
\mathbf{W}_{p}\bar{\mathbf{h}}^{p}
\right\|_{2}
},
\label{eq:clip_projection}
\end{equation}
where $\mathbf{W}_{c}$ and $\mathbf{W}_{p}$ are learnable projection
matrices.

Given a mini-batch of $B$ paired environmental and RM sequences, the
cross-modal similarity between the $i$-th environmental sequence and the
$j$-th RM sequence is defined as
\begin{equation}
S_{i,j}
=
\frac{
\left(
\mathbf{z}_{i}^{c}
\right)^{\top}
\mathbf{z}_{j}^{p}
}{
\tau_{\mathrm{clip}}
},
\label{eq:clip_similarity}
\end{equation}
where $\tau_{\mathrm{clip}}>0$ denotes the temperature parameter. The
CLIP module is optimized using the symmetric contrastive objective
\begin{equation}
\begin{aligned}
\mathcal{L}_{\mathrm{CLIP}}
=
-\frac{1}{2B}
\sum_{i=1}^{B}
\Bigg[
&
\log
\frac{
\exp(S_{i,i})
}{
\sum_{j=1}^{B}\exp(S_{i,j})
}
\\
+
&
\log
\frac{
\exp(S_{i,i})
}{
\sum_{j=1}^{B}\exp(S_{j,i})
}
\Bigg].
\end{aligned}
\label{eq:clip_loss}
\end{equation}

By distinguishing matched environment--propagation sequence pairs from
mismatched pairs, the environmental encoder is encouraged to retain
environmental variations that are relevant to propagation evolution,
rather than merely encoding geometric layouts.

After CLIP pretraining, the environmental encoder is
retained to generate the physics-semantic feature matrix
\begin{equation}
\mathbf{c}_{\hat{T}}
=
g_{\phi^{\star}}
\left(
\mathcal{C}_{1:\hat{T}}
\right),
\label{eq:physics_semantic_feature}
\end{equation}
where $\phi^{\star}$ denotes the propagation encoder parameters obtained
from CLIP pretraining. Therefore, $\mathbf{c}_{\hat{T}}$ summarizes the
historical propagation evolution while embedding the
environment--propagation correspondence learned from simulated data.

\begin{figure}[t]
\centering
\includegraphics[height=5.5cm,width=8.5cm]{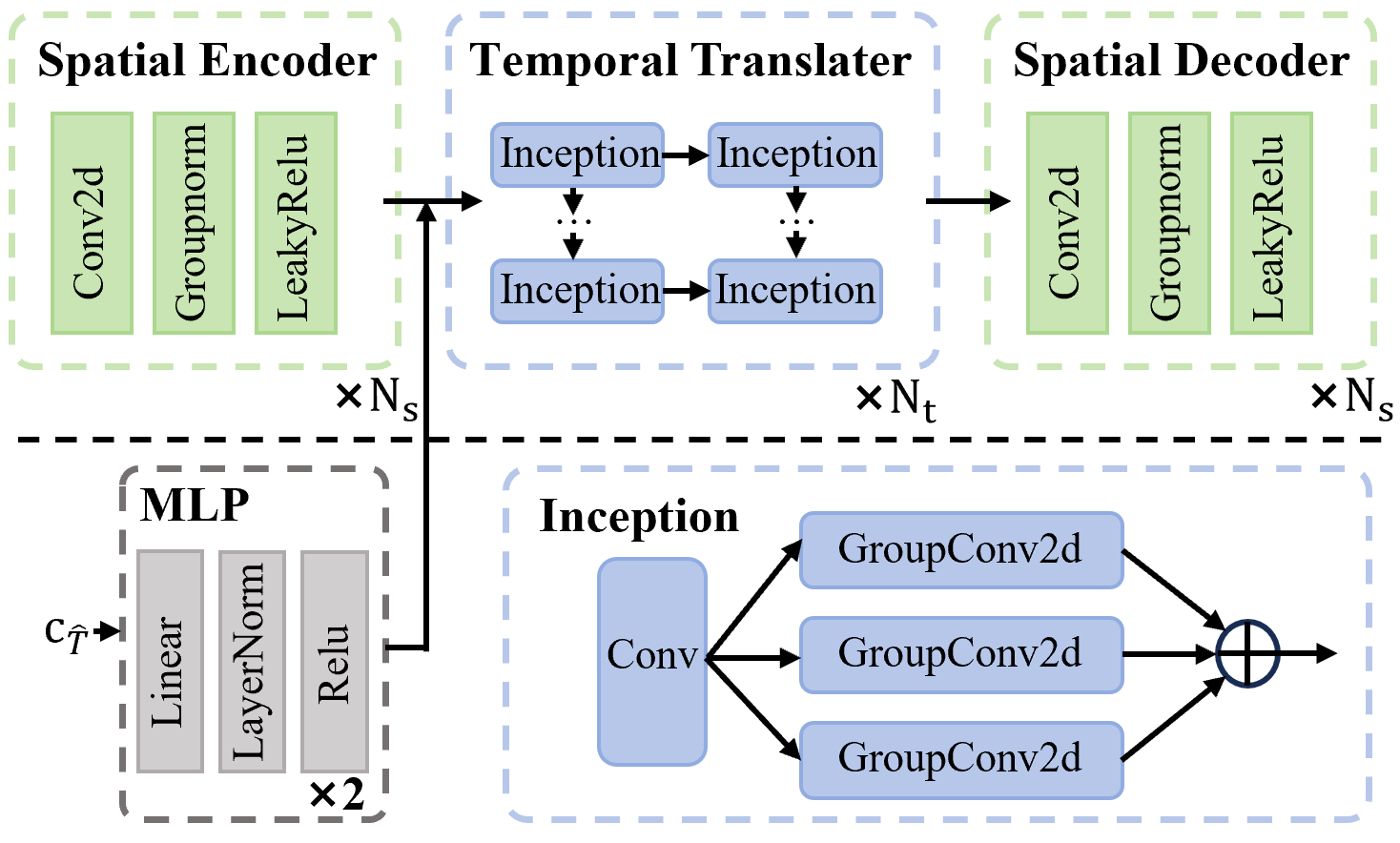}
\caption{Architecture of the condition-aware spatio-temporal backbone.
The physics-semantic condition $c_{\hat{T}}$ is projected and injected into the temporal translator, where multi-scale Inception blocks capture propagation dynamics for future RM prediction.}
\label{fig:backbone}
\end{figure}

\textit{2) Physics-Semantic-Guided RM Prediction:} The second stage of RadioCLIP predicts future RMs using both historical RM
observations and the learned physics-semantic feature
$\mathbf{c}_{\hat{T}}$. The RM input at time slot $t$ is represented
as
\begin{equation}
X_t
=
\operatorname{stack}
\left(
P_t,
M_t
\right),
\qquad
t=1,\ldots,\hat{T},
\label{eq:rm_input}
\end{equation}
where $M_t\in\{0,1\}^{N\times N}$ denotes the observation mask. During
simulation pretraining, complete RM observations are available and hence
\begin{equation}
M_t
=
\mathbf{1}_{N\times N},
\quad
t=1,\ldots,T.
\label{eq:sim_dense_mask}
\end{equation}

Let
\begin{equation}
\mathcal{X}_{1:\hat{T}}
\triangleq
\{X_t\}_{t=1}^{\hat{T}}
\end{equation}
denote the historical RM input sequence. As shown in Fig.~\ref{fig:backbone}, each $X_t$ is first processed
independently by a spatial encoder $\mathcal{E}_{\alpha}(\cdot)$ with parameters $\alpha$,
consisting of $N_s$ stacked convolutional blocks. Each block comprises a
2-D convolution, layer normalization, and a LeakyReLU activation\cite{gao2022simvp}. The
resulting latent RM features are expressed as
\begin{equation}
\mathbf{F}_{1:\hat{T}}
=
\mathcal{E}_{\alpha}
\left(
\mathcal{X}_{1:\hat{T}}
\right),
\label{eq:rm_spatial_encoding}
\end{equation}
where
$\mathbf{F}_{1:\hat{T}}
\in
\mathbb{R}^{\hat{T}\times C_f\times H_f\times W_f}$,
and $C_f$, $H_f$, and $W_f$ denote the channel dimension and spatial
resolution of the latent propagation features, respectively. This spatial
encoding compresses the historical RMs while preserving their dominant
propagation structures.

To incorporate the physics-semantic representation without directly
concatenating heterogeneous features, we introduce a lightweight
conditioning network $\mathcal{Q}_{\omega}(\cdot)$ with parameters $\omega$. Specifically,
$\mathcal{Q}_{\omega}(\cdot)$ is implemented using a two-layer
multilayer perceptron (MLP), which projects
$\mathbf{c}_{\hat{T}}$ into two $C_f$-dimensional vectors:
\begin{equation}
\begin{bmatrix}
\boldsymbol{\gamma}_{c} \\
\boldsymbol{\beta}_{c}
\end{bmatrix}
=
\mathcal{Q}_{\omega}
\left(
\mathbf{c}_{\hat{T}}
\right),
\label{eq:condition_projection}
\end{equation}
where
$\boldsymbol{\gamma}_{c},\boldsymbol{\beta}_{c}
\in\mathbb{R}^{C_f}$
denote channel-wise scaling and shifting coefficients, respectively.
The historical propagation features are then modulated as
\begin{equation}
\widetilde{\mathbf{F}}_{1:\hat{T}}
=
\left(
\mathbf{1}
+
\boldsymbol{\gamma}_{c}
\right)
\odot
\mathbf{F}_{1:\hat{T}}
+
\boldsymbol{\beta}_{c},
\label{eq:physics_semantic_modulation}
\end{equation}
where the modulation coefficients are broadcast over the time and spatial
dimensions. In this manner, the physics-semantic representation adjusts
the latent propagation features according to the environmental evolution
learned during cross-modal pretraining.

The conditioned features are subsequently processed by a temporal
evolution module $\mathcal{G}_{\beta}(\cdot)$ with parameters $\beta$. Specifically, the temporal
and channel dimensions of
$\widetilde{\mathbf{F}}_{1:\hat{T}}$ are first merged, after which
$N_t$ stacked multi-scale Inception blocks are employed to model the
evolution of the latent propagation field. Each Inception block contains
a $1\times1$ bottleneck convolution followed by multiple parallel grouped
convolutions with different receptive fields. The temporal evolution is
therefore expressed as
\begin{equation}
\mathbf{U}_{\hat{T}+1:T}
=
\mathcal{G}_{\beta}
\left(
\widetilde{\mathbf{F}}_{1:\hat{T}}
\right),
\label{eq:temporal_evolution}
\end{equation}
where
$\mathbf{U}_{\hat{T}+1:T}$ represents the predicted latent propagation
states for future time slots. The multi-scale convolutions allow the
module to jointly capture local propagation variations and broader spatial
dependencies while directly mapping historical latent states to future
ones without recurrent prediction.

Finally, a spatial decoder $\mathcal{D}_{\gamma}(\cdot)$ with parameters $\gamma$ reconstructs the
future latent states into full-resolution RMs. The decoder consists of
$N_s$ stacked upsampling convolutional blocks, each implemented using a
2-D transposed convolution followed by group normalization and a
LeakyReLU activation. Accordingly,
\begin{equation}
\widehat{\mathcal{P}}_{\hat{T}+1:T}
=
\mathcal{D}_{\gamma}
\left(
\mathbf{U}_{\hat{T}+1:T}
\right),
\label{eq:rm_reconstruction}
\end{equation}

The RM prediction network is optimized using
\begin{equation}
\mathcal{L}_{\mathrm{pred}}
=
\frac{1}{(T-\hat{T})N^{2}}
\sum_{t=\hat{T}+1}^{T}
\left\|
\widehat{P}_t-P_t
\right\|_{F}^{2},
\label{eq:prediction_loss}
\end{equation}
where $\|\cdot\|_{F}$ denotes the Frobenius norm. RadioCLIP is trained
sequentially: the cross-modal module is first pretrained using
$\mathcal{L}_{\mathrm{CLIP}}$, after which the environmental encoder is
fixed to generate $\mathbf{c}_{\hat{T}}$, and the RM prediction network
is optimized using $\mathcal{L}_{\mathrm{pred}}$.

\subsection{SparseUOT: Sparse-Observation-Aware Simulation-to-Real Adaptation}
\label{subsec:sparseuot}

Although RadioCLIP improves dynamic RM prediction by incorporating
propagation-relevant environmental semantics, directly deploying the
simulation-trained model in real environments remains challenging due to
two mismatches: the severe sparsity of real measurements and the
distribution discrepancy between simulated and measured propagation
fields. SparseUOT addresses these issues sequentially. Starting from the
RadioCLIP predictor pretrained on complete simulated RMs, we first perform
sparse-mask simulation refinement to expose the predictor to
deployment-like observation patterns. We then employ unbalanced optimal
transport (UOT) to selectively align compatible propagation
representations between the simulation and real domains under sparse
real-world supervision.

\textit{1) Sparse-Mask Simulation Refinement:} The simulation-trained RadioCLIP predictor initially observes complete RMs,
where the observation mask is given by
$M_t=\mathbf{1}_{N\times N}$. To bridge the observation-density mismatch
before real-domain adaptation, we randomly sparsify both the historical RM
inputs and the future RM supervision during simulation refinement.

For a simulated RM $P_t^{s}$, we construct a binary observation mask
\begin{equation}
M_t^{s}
=
[m_{i,j,t}^{s}]_{i,j=1}^{N}
\in
\{0,1\}^{N\times N},
\qquad
\left\|M_t^{s}\right\|_{0}
=
K_t^{s},
\label{eq:sim_sparse_mask}
\end{equation}
where $K_t^{s}$ denotes the number of retained receiver measurements at
time slot $t$. During training, $K_t^{s}$ is sampled from a
sparse-observation distribution $p_{K}$, and the corresponding receiver
locations are randomly resampled across time slots and training
iterations.

The sparsely observed RM is then given by
\begin{equation}
\widetilde{P}_t^{s}
=
M_t^{s}
\odot
P_t^{s},
\label{eq:sim_sparse_rm}
\end{equation}
where $\odot$ denotes element-wise multiplication. Following the input
representation introduced in Section~\ref{subsec:radioclip}, the sparse
RM value map and its observation mask are stacked along the channel
dimension:
\begin{equation}
X_t^{s}
=
\operatorname{stack}
\left(
\widetilde{P}_t^{s},
M_t^{s}
\right),
\qquad
t=1,\ldots,\hat{T}.
\label{eq:sim_sparse_input}
\end{equation}
Accordingly, the historical sparse input sequence is defined as
\begin{equation}
\mathcal{X}_{1:\hat{T}}^{s}
\triangleq
\{X_t^{s}\}_{t=1}^{\hat{T}}.
\label{eq:sim_sparse_sequence}
\end{equation}

\IncMargin{1em}
\begin{algorithm}[htbp]
\caption{Overall Training and Adaptation Procedure of DyTransRM}
\label{alg:dytransrm}

\KwIn{
Simulated dataset $\mathcal{D}_s$;
sparse real dataset $\mathcal{D}_r$;
sparse-observation distribution $p_K$;
$\lambda_{\mathrm{obs}}$ and $\lambda_{\mathrm{uot}}$
}

\KwOut{
Adapted RM predictor $\mathcal{F}_{\Theta^\star}$
}

Initialize $f_{\psi}$, $g_{\phi}$, projection heads
$\mathbf{W}_c$, $\mathbf{W}_p$, and predictor
$\mathcal{F}_{\Theta}$\;

\tcp{\textbf{Stage I: Cross-modal physics-semantic pretraining}}

\For{each simulated mini-batch}{
    Extract $\mathcal{C}_{1:\hat{T}}^{s}$ and $\mathcal{P}_{1:\hat{T}}^{s}$\;

    Compute
    $\mathbf{H}_{\hat{T}}^{c}
    \leftarrow
    f_{\psi}(\mathcal{C}_{1:\hat{T}}^{s})$
    and
    $\mathbf{H}_{\hat{T}}^{p}
    \leftarrow
    g_{\phi}(\mathcal{P}_{1:\hat{T}}^{s})$\;

    Obtain the normalized cross-modal embeddings
    $\mathbf{z}^{c}$ and $\mathbf{z}^{p}$\;

    Compute $\mathcal{L}_{\mathrm{CLIP}}$
    using~(\ref{eq:clip_loss})\;

    Update $\psi$, $\phi$, $\mathbf{W}_c$, and $\mathbf{W}_p$\;
}

Freeze the pretrained encoders as
$f_{\psi^\star}$ and $g_{\phi^\star}$\;

\tcp{\textbf{Stage II: RadioCLIP prediction training}}

\For{each simulated mini-batch}{
    Set $M_t^{s}\leftarrow\mathbf{1}_{N\times N}$,
    $t=1,\ldots,\hat{T}$\;

    Construct
    $X_t^{s}
    \leftarrow
    \operatorname{stack}(P_t^{s},M_t^{s})$\;

    Compute
    $\mathbf{c}_{\hat{T}}^{s}
    \leftarrow
    f_{\psi^\star}(\mathcal{C}_{1:\hat{T}}^{s})$\;

    Predict $\widehat{\mathcal{P}}_{\hat{T}+1:T}^{s}
    \leftarrow
    \mathcal{F}_{\Theta}(\mathcal{X}_{1:\hat{T}}^{s}, \mathbf{c}_{\hat{T}}^{s})$\;

    Update $\Theta$ by minimizing
    $\mathcal{L}_{\mathrm{pred}}$
    in~(\ref{eq:prediction_loss})\;
}

\tcp{\textbf{Stage III: Sparse-mask simulation refinement}}

\For{each simulated mini-batch}{
    Sample
    $\{M_t^{s}\}_{t=1}^{T}$
    according to $p_K$\;

    Construct
    $\widetilde{P}_t^{s}
    \leftarrow
    M_t^{s}\odot P_t^{s}$\;

    Construct historical sparse inputs
    $X_t^{s}
    \leftarrow
    \operatorname{stack}
    (\widetilde{P}_t^{s},M_t^{s})$\;

    Compute
    $\mathbf{c}_{\hat{T}}^{s}
    \leftarrow
    f_{\psi^\star}(\mathcal{C}_{1:\hat{T}}^{s})$
    and predict $\widehat{\mathcal{P}}_{\hat{T}+1:T}^{s}$\;

    Compute $\mathcal{L}_{\mathrm{sp}}$
    over
    $\{\Omega_t^{s}\}_{t=\hat{T}+1}^{T}$
    using~(\ref{eq:sim_sparse_loss})\;

    Update $\Theta$\;
}

\tcp{\textbf{Stage IV: UOT-based real-domain adaptation}}

\For{each adaptation iteration}{
    Sample one simulated mini-batch and one real mini-batch\;

    Construct $\mathcal{X}_{1:\hat{T}}^{s}$
    and
    $\mathcal{X}_{1:\hat{T}}^{r}$
    from their sparse RM observations and masks\;

    Forward both domains through the shared predictor to obtain $\mathbf{U}_{\hat{T}+1:T}^{s}$ and $\mathbf{U}_{\hat{T}+1:T}^{r}$\;

    Pool and normalize the latent propagation representations\;

    Compute $\mathcal{L}_{\mathrm{UOT}}$ and $\mathcal{L}_{\mathrm{obs}}$
    using~(\ref{eq:uot_loss}),(\ref{eq:real_observation_loss})\;

    Update $\Theta$ by minimizing $\mathcal{L}_{\mathrm{adapt}}$
    in~(\ref{eq:sparseuot_objective})\;
}

\Return{$\mathcal{F}_{\Theta^\star}$}\;

\end{algorithm}
\DecMargin{1em}

\begin{figure}[t]
\centering
\subfigure[]{
\begin{minipage}[c]{0.46\linewidth}
\centering
\includegraphics[height=3.6cm]{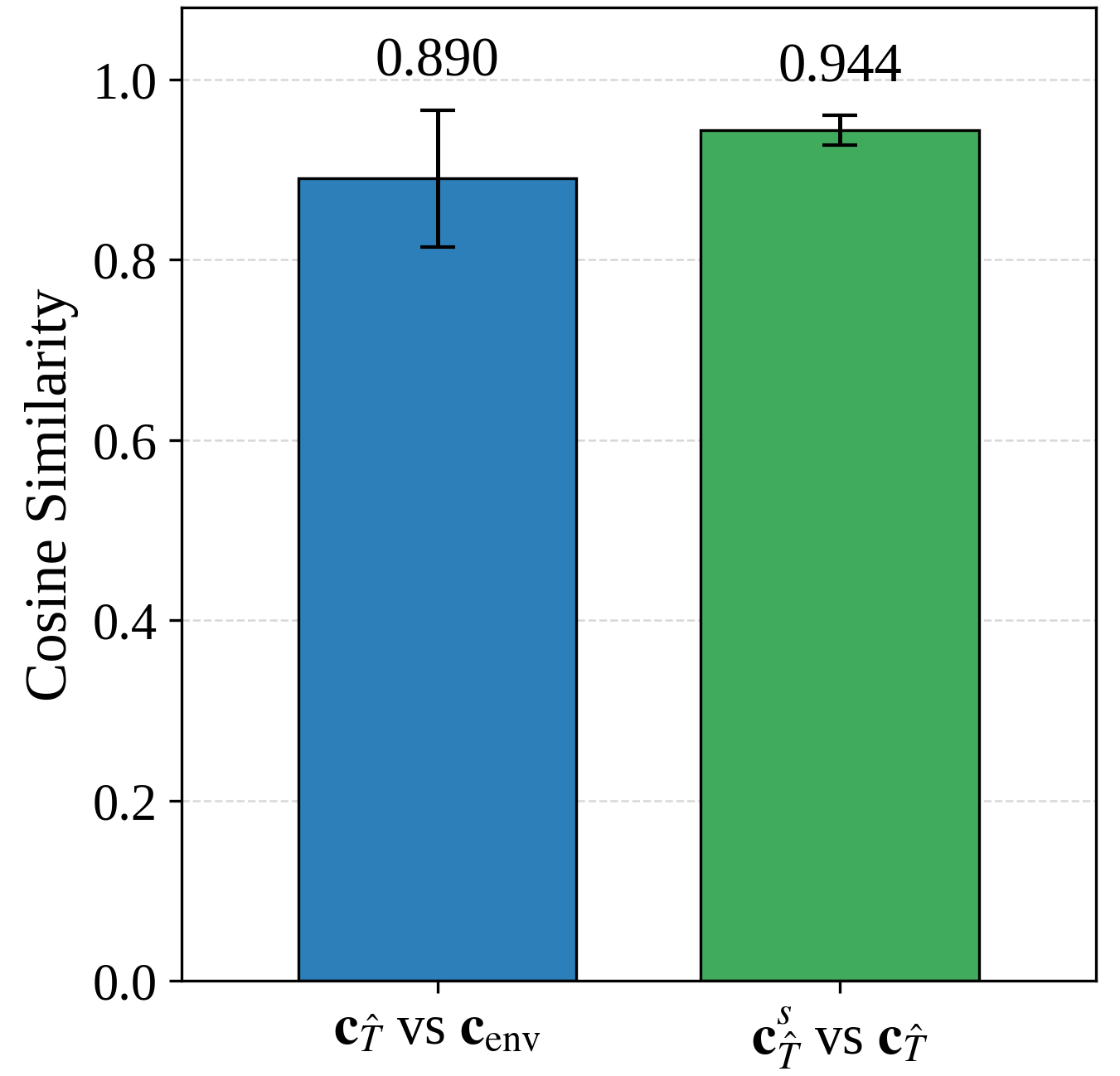}
\label{fig:feature_consistency}
\end{minipage}}
\hspace{0.1cm}
\subfigure[]{
\begin{minipage}[c]{0.46\linewidth}
\centering
\includegraphics[height=3.6cm]{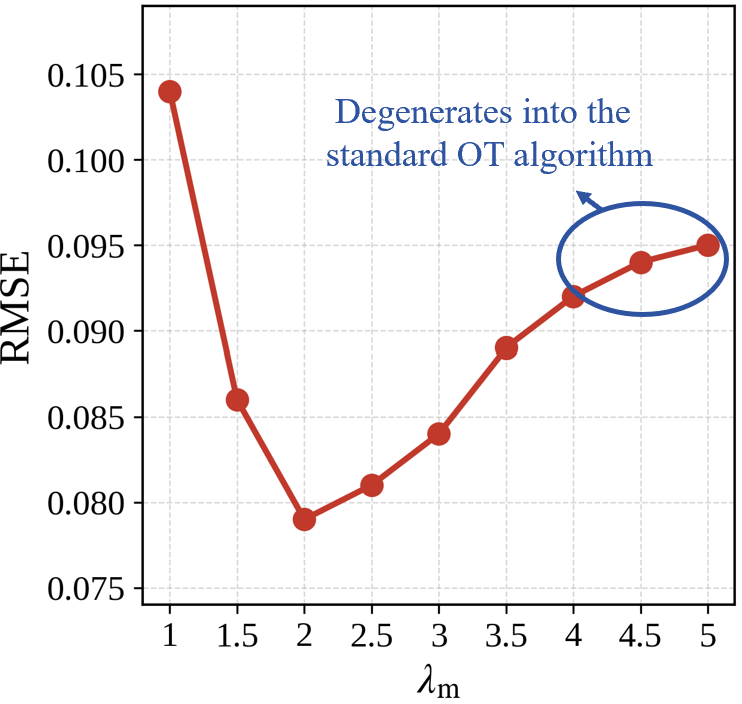}
\label{fig:OT}
\end{minipage}}
\caption{
Analysis of semantic consistency and UOT performance.
(a) Semantic feature consistency between full radio map observations and sparse measurements. (b) Effect of the marginal relaxation coefficient $\lambda_m$ in unbalanced optimal transport on the real-domain prediction performance.}
\label{fig:performance}
\end{figure}

The pretrained environmental encoder remains fixed and generates the
physics-semantic representation
\begin{equation}
\mathbf{c}_{\hat{T}}^{s}
=
g_{\phi^{\star}}
\left(
\mathcal{C}_{1:\hat{T}}^{s}
\right).
\label{eq:sim_physics_semantic_feature}
\end{equation}
The future RM sequence is then predicted by the same RadioCLIP prediction
network:
\begin{equation}
\widehat{\mathcal{P}}_{\hat{T}+1:T}^{s}
=
\mathcal{F}_{\Theta}
\left(
\mathcal{X}_{1:\hat{T}}^{s},
\mathbf{c}_{\hat{T}}^{s}
\right),
\label{eq:sim_sparse_prediction}
\end{equation}
where
$\Theta=\{\alpha,\omega,\beta,\gamma\}$
collects the parameters of the spatial encoder, conditioning network,
temporal evolution module, and spatial decoder. As shown in Fig.~\ref{fig:feature_consistency}, the representation
obtained from sparse measurements achieves high cosine similarity with the reference semantic representation, which demonstrates
that although only sparse propagation measurements
$\tilde{P}_{1:\hat{T}}^{r}$ are available during deployment, the pretrained
encoder $g_{\phi^*}$ can still extract reliable semantic features
$c_{\hat{T}}^{r}$.

Different from the initial RadioCLIP training stage, the future simulated
RMs are also sparsely supervised. For each future time slot, we define the
set of retained receiver locations as
\begin{equation}
\Omega_t^{s}
\triangleq
\left\{
(i,j)
\,\middle|\,
m_{i,j,t}^{s}=1
\right\},
\qquad
t=\hat{T}+1,\ldots,T.
\label{eq:sim_observation_set}
\end{equation}
The sparse-mask simulation loss is therefore formulated as
\begin{equation}
\mathcal{L}_{\mathrm{sp}}
=
\frac{1}{T-\hat{T}}
\sum_{t=\hat{T}+1}^{T}
\frac{1}{|\Omega_t^{s}|}
\sum_{(i,j)\in\Omega_t^{s}}
\ell
\left(
\widehat{p}_{i,j,t}^{s},
p_{i,j,t}^{s}
\right),
\label{eq:sim_sparse_loss}
\end{equation}
where $\ell(\cdot,\cdot)$ denotes the point-wise prediction loss. In our
implementation, $\ell(a,b)=(a-b)^2$ is adopted to remain consistent with
the MSE objective used during RadioCLIP pretraining.

Although supervision is imposed only at the retained receiver locations,
the predictor continues to output a complete RM for every future time
slot. Moreover, continuously resampling the masks prevents the model from
memorizing a fixed receiver configuration and encourages it to infer the
underlying propagation field from different sparse observation patterns. 

\textit{2) Real-Domain Adaptation via Unbalanced Optimal Transport:} After sparse-mask simulation refinement, the predictor becomes compatible
with the observation regime of the real dataset. However, the propagation
distributions of the two domains remain different because of imperfect
environmental modeling and unmodeled real-world propagation effects.
We therefore further adapt the prediction network by aligning its latent
propagation representations between the simulation and real domains.

At time slot $t$, the available real RM measurements are represented by a
binary observation mask
\begin{equation}
M_t^{r}
=
[m_{i,j,t}^{r}]_{i,j=1}^{N}
\in
\{0,1\}^{N\times N},
\label{eq:real_sparse_mask}
\end{equation}
and the corresponding sparse RM is
\begin{equation}
\widetilde{P}_t^{r}
=
M_t^{r}
\odot
P_t^{r}.
\label{eq:real_sparse_rm}
\end{equation}
The real-domain predictor input follows exactly the same representation as
that used during sparse-mask simulation refinement:
\begin{equation}
X_t^{r}
=
\operatorname{stack}
\left(
\widetilde{P}_t^{r},
M_t^{r}
\right),
\qquad
t=1,\ldots,\hat{T},
\label{eq:real_sparse_input}
\end{equation}
with $\mathcal{X}_{1:\hat{T}}^{r}
\triangleq
\{X_t^{r}\}_{t=1}^{\hat{T}}$. 
Meanwhile, the fixed propagation encoder produces $\mathbf{c}_{\hat{T}}^{r}
=
g_{\phi^{\star}}
\left(
\widetilde{P}_{1:\hat{T}}^{r}
\right)$.

For each simulation or real sample, the prediction network generates the
future latent propagation states
$\mathbf{U}_{\hat{T}+1:T}^{d}$,
where $d\in\{s,r\}$ denotes the domain. Since these representations are
produced by the temporal evolution module immediately before RM
reconstruction, they characterize the predicted propagation evolution in a
common latent space. For the $b$-th sample from domain $d$, we obtain a
compact representation through temporal-spatial pooling and
$\ell_2$ normalization:
\begin{equation}
\overline{\mathbf{h}}_{b}^{d}
=
\operatorname{Pool}
\left(
\mathbf{U}_{\hat{T}+1:T}^{d,(b)}
\right),
\label{eq:uot_pooling}
\end{equation}
\begin{equation}
\mathbf{h}_{b}^{d}
=
\frac{
\overline{\mathbf{h}}_{b}^{d}
}{
\left\|
\overline{\mathbf{h}}_{b}^{d}
\right\|_{2}
},
\qquad
d\in\{s,r\},
\label{eq:normalized_uot_features}
\end{equation}
where $\operatorname{Pool}(\cdot)$ denotes average pooling over the
prediction time and spatial dimensions.

Given a simulation mini-batch of size $B_s$ and a real mini-batch of size
$B_r$, the transport cost between the $i$-th simulation sample and the
$j$-th real sample is defined as
\begin{equation}
C_{i,j}
=
\left\|
\mathbf{h}_{i}^{s}
-
\mathbf{h}_{j}^{r}
\right\|_{2}^{2}.
\label{eq:uot_cost}
\end{equation}
Instead of enforcing one-to-one or globally balanced matching, UOT allows
only compatible portions of the two mini-batch distributions to be
transported. This property is particularly useful when the simulated
dataset spans a broader propagation distribution than the limited real
measurements.

Let
\begin{equation}
\mathbf{a}
=
\frac{1}{B_s}\mathbf{1}_{B_s},
\qquad
\mathbf{b}
=
\frac{1}{B_r}\mathbf{1}_{B_r}
\end{equation}
denote the empirical sample weights of the simulation and real
mini-batches, respectively. The UOT transport plan is obtained by
\begin{equation}
\begin{aligned}
\boldsymbol{\Pi}^{\star}
=
\arg\min_{\boldsymbol{\Pi}\geq 0}
\;&
\left\langle
\boldsymbol{\Pi},
\mathbf{C}
\right\rangle_{F}
+
\varepsilon_{\mathrm{ot}}
\Omega
\left(
\boldsymbol{\Pi}
\right)
\\
&+
\lambda_{\mathrm{m}}
\mathrm{KL}
\left(
\boldsymbol{\Pi}\mathbf{1}_{B_r}
\|
\mathbf{a}
\right)
\\
&+
\lambda_{\mathrm{m}}
\mathrm{KL}
\left(
\boldsymbol{\Pi}^{\top}\mathbf{1}_{B_s}
\|
\mathbf{b}
\right),
\end{aligned}
\label{eq:uot_plan}
\end{equation}
where $\varepsilon_{\mathrm{ot}}$ controls the entropic regularization,
$\lambda_{\mathrm{m}}$ controls the relaxation of the marginal
constraints, and
\begin{equation}
\Omega
\left(
\boldsymbol{\Pi}
\right)
=
\sum_{i=1}^{B_s}
\sum_{j=1}^{B_r}
\Pi_{i,j}
\left(
\log\Pi_{i,j}-1
\right)
\label{eq:uot_entropy}
\end{equation}
denotes the entropic regularization term.

The resulting transport plan is used to construct the cross-domain
representation-alignment loss
\begin{equation}
\mathcal{L}_{\mathrm{UOT}}
=
\left\langle
\operatorname{sg}
\left(
\boldsymbol{\Pi}^{\star}
\right),
\mathbf{C}
\right\rangle_{F},
\label{eq:uot_loss}
\end{equation}
where $\operatorname{sg}(\cdot)$ denotes the stop-gradient operation.
Hence, the transport plan determines which simulation-real pairs should
contribute to domain alignment, while gradients are propagated through
their latent propagation representations. As shown in
Fig.~\ref{fig:OT}, when $\lambda_m$ becomes excessively large, the performance
gradually degrades. This is because a strong marginal constraint forces the
unbalanced optimal transport process to approach the behavior of standard
balanced optimal transport, where all samples are required to be matched
regardless of their compatibility. Such forced alignment may introduce
incorrect correspondences between simulation and real propagation
distributions, thereby reducing adaptation effectiveness.

In addition to representation alignment, the available future real
measurements provide direct prediction supervision. Define
\begin{equation}
\Omega_t^{r}
\triangleq
\left\{
(i,j)
\,\middle|\,
m_{i,j,t}^{r}=1
\right\},
\qquad
t=\hat{T}+1,\ldots,T.
\label{eq:real_observation_set}
\end{equation}
The real-domain observation loss is evaluated only at these measured
receiver locations:
\begin{equation}
\mathcal{L}_{\mathrm{obs}}
=
\frac{1}{T-\hat{T}}
\sum_{t=\hat{T}+1}^{T}
\frac{1}{|\Omega_t^{r}|}
\sum_{(i,j)\in\Omega_t^{r}}
\rho
\left(
\widehat{p}_{i,j,t}^{r}
-
p_{i,j,t}^{r}
\right),
\label{eq:real_observation_loss}
\end{equation}
where $\rho(\cdot)$ denotes the Huber penalty. Therefore, spatially
complete real RMs are not required during adaptation.

The final SparseUOT adaptation objective is
\begin{equation}
\mathcal{L}_{\mathrm{adapt}}
=
\lambda_{\mathrm{obs}}
\mathcal{L}_{\mathrm{obs}}
+
\lambda_{\mathrm{uot}}
\mathcal{L}_{\mathrm{UOT}},
\label{eq:sparseuot_objective}
\end{equation}
where $\lambda_{\mathrm{obs}}$ and $\lambda_{\mathrm{uot}}$ balance
sparse real-domain supervision and cross-domain representation alignment,
respectively. The entire algorithm follows a four-stage architecture, as illustrated in Algorithm~\ref{alg:dytransrm}.

\section{Numerical Results}
\label{sect:num_res}

\begin{table*}[t]
\centering
\renewcommand{\arraystretch}{1.2} % 增大行高，视觉更丰满
\large % 整体放大字体
\caption{Dynamic RM Prediction Performance on the RadioMapMotion Dataset.
Lower NMSE and RMSE are better, while higher SSIM and PSNR are better.}
\label{tab:quantitative_comparison}
\begin{tabular}{|c|c|c|c|c|c|c}
\hline
\textbf{Metric} & \textbf{CSI-MAE~\cite{jiang2026csi}} & \textbf{JEPA-CFM~\cite{gao2026jepa}} & \textbf{BERT\cite{zhao2025csi}} & \textbf{ConvLSTM\cite{cheng2026radiomapmotion}} & \textbf{RadioCLIP} \\
\hline
NMSE        & 0.008 & 0.011 & 0.0079 & 0.0025 & \textbf{0.0018} \\
RMSE        & 0.028 & 0.031 & 0.026 & 0.015 & \textbf{0.012} \\
SSIM $\uparrow$ & 0.9128 & 0.9437 & 0.9119 & 0.9779 & \textbf{0.9824} \\
PSNR $\uparrow$ & 31.01 & 30.10 & 31.62 & 36.37 & \textbf{39.13} \\
\hline
\end{tabular}
\end{table*}

This section evaluates DyTransRM from two complementary perspectives:
simulation-domain dynamic RM prediction and simulation-to-real adaptation
under sparse real-world measurements. We first introduce the datasets,
evaluation protocol, implementation details, and comparison baselines,
followed by quantitative and qualitative performance analyses.

\begin{figure*}[t] %[h]
\centering  
\includegraphics[height=6.5cm,width=18cm]{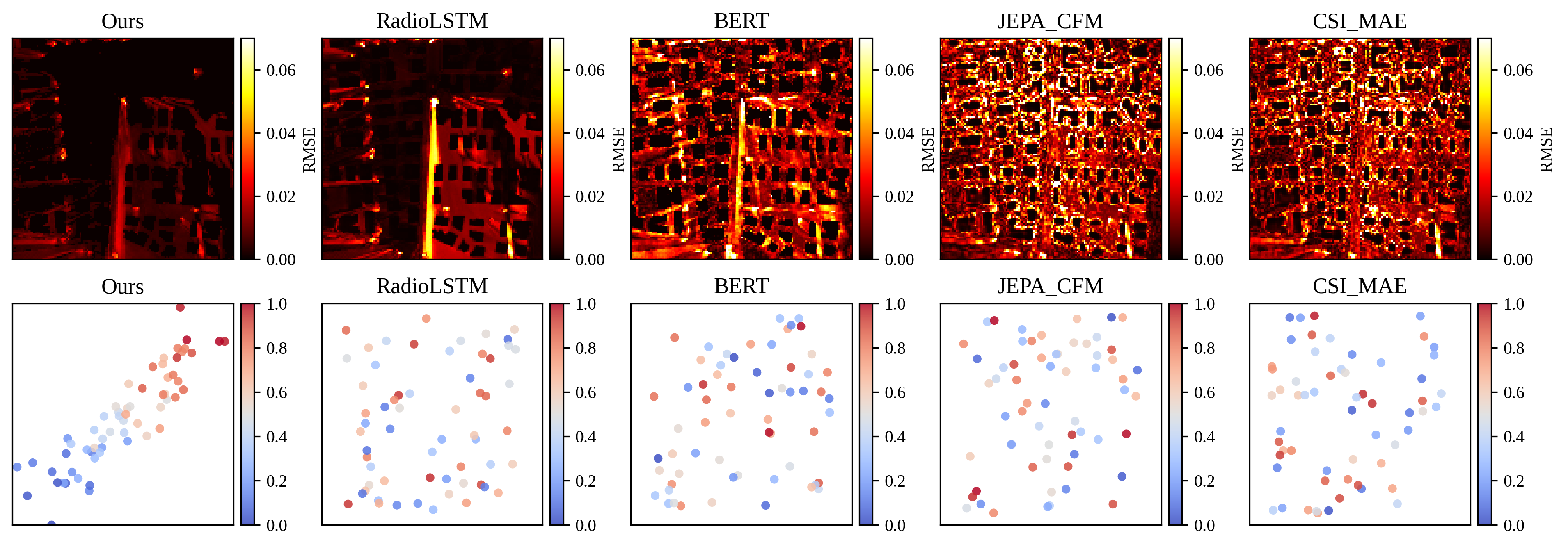}
\caption{The error distribution and the t-SNE visualization of conditional embeddings. The first row illustrates the error distribution of the generated results on the target time slot, while each point in the second row denotes a test sample after projecting the condition embedding fed into the temporal evolition module onto a two-dimensional space. }
\label{fig:tsne}
\end{figure*}

\subsection{Simulation Settings}
\label{subsec:experimental_setup}

\textit{1) Datasets}

We evaluate DyTransRM using a large-scale simulated dynamic RM dataset and two target-domain datasets. The simulated dataset is used to learn dynamic propagation knowledge and evaluate simulation-domain prediction accuracy, whereas the target-domain datasets are used to assess sparse simulation-to-real adaptation and robustness to different measurement densities.

\textit{RadioMapMotion~\cite{cheng2026radiomapmotion}:}
We adopt the RadioMapMotion dataset for simulation-domain training and evaluation. RadioMapMotion contains temporally continuous RM sequences generated from dynamic urban environments with moving vehicles. The dataset comprises 300 urban environments, five vehicle trajectories per environment, and 20 fixed transmitter locations. Each environment--trajectory--transmitter combination forms a 15-time-slot RM sequence, resulting in 30,000 sequences and 450,000 individual RMs. Each RM has a spatial resolution of $256\times256$, corresponding to a $256~\mathrm{m}\times256~\mathrm{m}$ area with a horizontal resolution of $1~\mathrm{m}$. The propagation data are generated using the Dominant Path Model at a carrier frequency of $3.5~\mathrm{GHz}$, with a transmit power of $23~\mathrm{dBm}$ and transmitter/receiver heights of $1.5~\mathrm{m}$. Adjacent time slots are separated by $0.1~\mathrm{s}$, such that the dataset explicitly preserves propagation-field evolution induced by vehicle motion.

Following the prediction protocol of RadioMapMotion, each sequence is divided into $\hat T=10$ historical time slots and $T-\hat T=5$ future time slots. We follow the trajectory-disjoint partition adopted by RadioMapMotion: the first three trajectories from environments 0--249 are used for training, the fourth trajectory is used for validation, and the fifth trajectory is reserved for testing. Therefore, the reported simulation-domain performance evaluates prediction on vehicle trajectories that are not observed during model training. The same partition, historical sequence length, and prediction horizon are used for all comparison methods.

\textit{Vienna\cite{wiedner2026vienna}:}
For the primary target-domain evaluation, we employ the Vienna 4G/5G Drive-Test Dataset, which contains georeferenced and temporally indexed cellular measurements collected from operational LTE and 5G NR networks in Vienna, Austria. Unlike RadioMapMotion, the Vienna dataset provides irregularly distributed receiver measurements rather than spatially complete RM sequences. To maintain consistency with the propagation quantity used in the simulation domain, we adopt the RSRP field from the passive LTE scanner measurements of Operator~A. For each measurement, the serving cell is defined as the cell with the strongest RSRP, and measurements associated with different serving cells are processed separately to avoid mixing propagation fields generated by different transmitters. We retain the top $200$ cells ranked by the number of serving measurements, each contributing at least $100$ measurement points.

The geographic coordinates are transformed from WGS84 to the metric EPSG:31287 projection. Measurements are grouped into non-overlapping local regions of $512\,\mathrm{m} \times 512\,\mathrm{m}$ and projected onto a $256 \times 256$ spatial grid, corresponding to a resolution of approximately $2$\,m per pixel. When multiple measurements fall into the same grid cell within one time slot, their RSRP values are averaged. For each (cell, region) pair containing at least $75$ measurements, the chronologically sorted measurements are evenly partitioned into fifteen consecutive slots with at least five points per slot. The first $\hat{T}=10$ slots form the historical observation interval and the remaining $T-\hat{T}=5$ slots constitute the prediction horizon, consistent with the simulation-domain protocol. RSRP values are clipped to $[-135, -39.5]$\,dBm and linearly normalized to $[0, 1]$, matching the value range of the simulation domain.

Importantly, only grid cells containing genuine drive-test measurements are retained as target-domain observations, while unmeasured cells are neither spatially interpolated nor treated as pseudo ground truth. To prevent temporal leakage between adaptation and evaluation, the sequences are partitioned into non-overlapping adaptation, validation, and test subsets with proportions of $70\%$, $15\%$, and $15\%$, respectively.

\textit{WiWorld\cite{jiao2026wiworld}:}
As a secondary target-domain benchmark, we employ the WiWorld-RealData dataset, a real-world outdoor multi-modal channel dataset collected along mobile routes on the campus of Beijing University of Posts and Telecommunications. The measurements were conducted with a fixed transmitter and a mobile receiver. The current public release provides one representative dual-band route with measured channel impulse responses~(CIRs) at both $3.7$\,GHz and $6.775$\,GHz. To maintain consistency with the carrier-frequency band used in the simulation domain, we select the $3.7$\,GHz band and derive the path-loss field from the measured CIRs. We follow the same temporal prediction formulation and sparse-observation representation as those used for the Vienna dataset. Specifically, only the receiver locations selected by the corresponding observation mask are provided to the predictor, and the remaining locations are treated as unobserved rather than being reconstructed through preprocessing. The number of sampled receiver locations per time slot is varied consistently across the two target datasets, allowing the robustness of the proposed adaptation mechanism to different measurement budgets to be examined under a common protocol.

\textit{2) Evaluation Protocol and Metrics:}

Different evaluation metrics are adopted for the simulation and real domains according to the availability of reference measurements. For the simulation-domain experiment, spatially complete ground-truth RMs are available. We therefore evaluate the prediction results using four metrics, namely NMSE, RMSE, SSIM, and PSNR. NMSE and RMSE measure the point-wise prediction error, whereas SSIM and PSNR evaluate the structural fidelity of the predicted propagation fields. Lower NMSE and RMSE indicate better prediction accuracy, while higher SSIM and PSNR indicate better reconstruction quality.

Specifically, the simulation-domain NMSE is defined as
\begin{equation}
\mathrm{NMSE}
=
\frac{
\displaystyle
\sum_{t=\hat T+1}^{T}
\left\|
\hat{\mathbf P}_{t}-\mathbf P_{t}
\right\|_{F}^{2}
}{
\displaystyle
\sum_{t=\hat T+1}^{T}
\left\|
\mathbf P_{t}
\right\|_{F}^{2}
},
\label{eq:eval_nmse_sim}
\end{equation}
and the RMSE is given by
\begin{equation}
\mathrm{RMSE}
=
\sqrt{
\frac{
\displaystyle
\sum_{t=\hat T+1}^{T}
\left\|
\hat{\mathbf P}_{t}-\mathbf P_{t}
\right\|_{F}^{2}
}{
(T-\hat T)N^{2}
}
}.
\label{eq:eval_rmse_sim}
\end{equation}

The structural similarity index is computed as
\begin{equation}
\mathrm{SSIM}
=
\frac{1}{T-\hat T}
\sum_{t=\hat T+1}^{T}
\mathrm{SSIM}
\left(
\hat{\mathbf P}_{t},
\mathbf P_{t}
\right),
\label{eq:eval_ssim}
\end{equation}
while PSNR is defined as
\begin{equation}
\mathrm{PSNR}
=
\frac{1}{T-\hat T}
\sum_{t=\hat T+1}^{T}
10\log_{10}
\left(
\frac{P_{\max}^{2}}
{
\mathrm{MSE}
\left(
\hat{\mathbf P}_{t},
\mathbf P_{t}
\right)
}
\right),
\label{eq:eval_psnr}
\end{equation}
where $P_{\max}$ denotes the maximum value of the adopted RM representation.

For the real-domain experiments, spatially complete reference RMs are unavailable. Performance is therefore evaluated exclusively at genuinely measured receiver locations. Let $\Omega_{t}^{r,\mathrm{test}}$
denote the set of held-out real measurements available for evaluation at future time slot $t$. The real-domain NMSE is computed as
\begin{equation}
\mathrm{NMSE}_{r}
=
\frac{
\displaystyle
\sum_{t=\hat T+1}^{T}
\sum_{(i,j)\in\Omega_{t}^{r,\mathrm{test}}}
\left(
\hat p_{i,j,t}^{r}-p_{i,j,t}^{r}
\right)^2
}{
\displaystyle
\sum_{t=\hat T+1}^{T}
\sum_{(i,j)\in\Omega_{t}^{r,\mathrm{test}}}
\left(
p_{i,j,t}^{r}
\right)^2
},
\label{eq:eval_nmse_real}
\end{equation}
the RMSE is
\begin{equation}
\mathrm{RMSE}_{r}
=
\sqrt{
\frac{
\displaystyle
\sum_{t=\hat T+1}^{T}
\sum_{(i,j)\in\Omega_{t}^{r,\mathrm{test}}}
\left(
\hat p_{i,j,t}^{r}-p_{i,j,t}^{r}
\right)^2
}{
\displaystyle
\sum_{t=\hat T+1}^{T}
\left|
\Omega_{t}^{r,\mathrm{test}}
\right|
}
},
\label{eq:eval_rmse_real}
\end{equation}
and the MAE is
\begin{equation}
\mathrm{MAE}_{r}
=
\frac{
\displaystyle
\sum_{t=\hat T+1}^{T}
\sum_{(i,j)\in\Omega_{t}^{r,\mathrm{test}}}
\left|
\hat p_{i,j,t}^{r}-p_{i,j,t}^{r}
\right|
}{
\displaystyle
\sum_{t=\hat T+1}^{T}
\left|
\Omega_{t}^{r,\mathrm{test}}
\right|
}.
\label{eq:eval_mae_real}
\end{equation}

SSIM and PSNR are not reported for the real-domain experiments because the available measurements do not provide spatially complete reference RMs. Constructing dense reference maps through spatial interpolation would introduce additional assumptions into the evaluation and would be inconsistent with the practical setting considered in this work. Therefore, both real-domain adaptation and performance evaluation rely exclusively on genuinely measured receiver samples, without constructing interpolation-based pseudo ground truth.

\textit{3) Implementation Details:}
All methods are implemented in PyTorch~2.1.0 under Python~3.9 and trained
on a single NVIDIA GeForce RTX~4090 GPU. Unless otherwise specified,
identical RM preprocessing, historical sequence length, prediction horizon,
and dataset partitions are used for all comparison methods. Hyperparameters
are selected using the corresponding validation split and kept fixed during
test-set evaluation. The major architecture, training, and adaptation
parameters of DyTransRM are summarized in
Table~\ref{tab:simu_para}.

\begin{table} [t] % [h] 
\centering 
\caption{Implementation Parameters of DyTransRM.}  
\label{tab:simu_para}
\footnotesize
\begin{tabular}{c | c }  

\toprule 
RM grid size, $N\times N$
& $256\times256$
\\ 

\midrule
Sequence length, $T$
& 15
\\

\midrule
Historical sequence length, $\hat{T}$9p
& 10
\\

\midrule
Prediction horizon, $T-\hat{T}$
& 5
\\

\midrule
Spatial encoder/decoder blocks, $N_s$
& 4
\\ 

\midrule
Latent feature channels, $C_f$
& 64
\\

\midrule
Latent feature height, $H_f$
& 16
\\

\midrule
Latent feature width, $W_f$
& 16
\\

\midrule
Temporal Inception blocks, $N_t$
& 4
\\ 

\midrule
CLIP temperature, $\tau_{\mathrm{clip}}$
& 0.07
\\ 

\midrule
Stage~I batch size, $B$
& 16
\\ 

\midrule
Stage~I optimizer / learning rate
& AdamW / $1\times10^{-4}$
\\ 

\midrule
Stage~I training epochs
& 100
\\ 

\midrule
Stage~II batch size
& 8
\\ 

\midrule
Stage~II optimizer / learning rate
& AdamW / $1\times10^{-4}$
\\

\midrule
Stage~II training epochs
& 100
\\ 

\midrule
Sparse receiver count, $K_t^s\sim p_K$
& $p_K=\mathcal{U}\{5,\ldots,35\}$
\\

\midrule
Stage~III batch size
& 8
\\

\midrule
Stage~III optimizer / learning rate
& AdamW / $5\times10^{-5}$
\\

\midrule
Stage~III refinement epochs
& 30
\\ 

\midrule
Simulation/real batch sizes, $B_s/B_r$
& 8 / 8
\\ 

\midrule
Stage~IV optimizer / learning rate
& AdamW / $1\times10^{-5}$
\\ 

\midrule
Stage~IV adaptation iterations
& 2000
\\ 

\midrule
UOT entropic coefficient, $\varepsilon_{\mathrm{ot}}$
& 0.05
\\ 

\midrule
UOT marginal coefficient, $\lambda_{\mathrm{m}}$
& 2.0
\\ 

\midrule
Observation-loss weight, $\lambda_{\mathrm{obs}}$
& 1.0
\\ 

\midrule
UOT-loss weight, $\lambda_{\mathrm{uot}}$
& 0.1
\\ 

\bottomrule
\end{tabular}
\end{table}

\textit{4) Baselines:}

We compare DyTransRM with two groups of methods corresponding to the two stages of the proposed framework.

\textit{Simulation-domain prediction baselines:}
Four representative prediction and representation-learning methods are considered. RadioLSTM~\cite{cheng2026radiomapmotion} is a ConvLSTM--UNet-based sequence-to-sequence model specifically developed for dynamic RM forecasting and serves as the primary RM-specific spatio-temporal baseline. BERT~\cite{zhao2025csi} represents a Transformer-based sequence modeling paradigm and is adapted to predict future RM sequences from historical observations. CSI-MAE~\cite{jiang2026csi} employs masked reconstruction to learn transferable channel representations, whereas JEPA-CFM~\cite{gao2026jepa} adopts joint-embedding predictive learning to model high-level channel representations without relying exclusively on raw-value reconstruction. Since CSI-MAE and JEPA-CFM were originally designed for channel representation learning rather than multi-step dynamic RM forecasting, their input and output interfaces are adapted to the present prediction setting while retaining their respective core learning mechanisms. 

\textit{Simulation-to-real adaptation baselines:}
To evaluate SparseUOT under limited target-domain supervision, we compare against three representative adaptation strategies. LoRA~\cite{hu2021lora} freezes the simulation-pretrained backbone and introduces trainable low-rank parameters for parameter-efficient target-domain adaptation. MAML~\cite{muruganandham2025smart} employs meta-learning to obtain a model initialization that can be rapidly adapted using a limited number of target-domain measurements. Adversarial Learning~\cite{lu2026transfer} introduces a domain discriminator to encourage domain-invariant feature learning between simulated and measured propagation data.

\begin{table}[t]
\centering
\caption{Zero-Shot Simulation-to-Real Transfer Performance Without
Real-Domain Parameter Adaptation.}
\label{tab:zero_shot_transfer}
\normalsize  % 换成了标准字号，和正文大小一致
\setlength{\tabcolsep}{6pt} % 列间距微调回 6pt，保持紧凑但不过分拥挤
\renewcommand{\arraystretch}{1.20} % 行高微调，保持清爽
\begin{tabular}{lccc}
\toprule
\textbf{Method}
& \textbf{NMSE} $\downarrow$
& \textbf{RMSE} $\downarrow$
& \textbf{MAE} $\downarrow$ \\
\midrule
RadioLSTM~\cite{cheng2026radiomapmotion}
& 0.053 & 0.154 & 0.121 \\
BERT~\cite{zhao2025csi}
& 0.059 & 0.166 & 0.127 \\
CSI-MAE~\cite{jiang2026csi}
& 0.061 & 0.171 & 0.133 \\
JEPA-CFM~\cite{gao2026jepa}
& 0.064 & 0.175 & 0.138 \\
\midrule
\textbf{RadioCLIP (Ours)}
& \textbf{0.048} & \textbf{0.142} & \textbf{0.112} \\
\bottomrule
\end{tabular}
\end{table}

\subsection{Simulation-Domain Dynamic RM Prediction}
\label{subsec:simdomain}

Table~\ref{tab:quantitative_comparison} summarizes the prediction results
on the RadioMapMotion simulation dataset. RadioCLIP achieves the best
performance across all four metrics. Compared with the strongest baseline,
it reduces NMSE and RMSE by 28.0\% and 20.0\%, respectively, while improving
SSIM from 0.9779 to 0.9824 and PSNR by 2.76~dB. These results demonstrate
that explicitly incorporating environment-aware propagation semantics is
beneficial for dynamic RM prediction.

Fig.~\ref{fig:tsne} further provides insight into the source of this
improvement. The first row shows the spatial distribution of the prediction
error at the target time slot. The baseline methods exhibit pronounced
errors in regions whose propagation conditions change substantially with the
dynamic environment. Specifically, changes in the building configuration can
alter line-of-sight blockage as well as the geometry of reflected and
diffracted propagation paths, thereby producing localized and sometimes
abrupt variations in the RM. Such variations are difficult to infer solely
from temporal correlations among historical RM values. In contrast,
RadioCLIP explicitly associates environmental evolution with the
corresponding propagation evolution during cross-modal pretraining, resulting
in substantially smaller errors in these propagation-sensitive regions.

The second row of Fig.~\ref{fig:tsne} visualizes the learned condition
embeddings using t-SNE. The color of each point denotes the normalized RM
variation induced by changes in the dynamic building configuration, i.e.,
the degree to which the propagation field changes as the environment evolves.
A small value corresponds to a relatively stable propagation state, whereas
a large value indicates stronger RM variations caused by changes in blockage,
reflection, and diffraction conditions. The embeddings learned by RadioCLIP
exhibit a clear organization with respect to this propagation-variation
indicator, while those of the baseline methods are substantially more
scattered. Although t-SNE provides only a qualitative visualization, this
structure suggests that RadioCLIP preserves environment-dependent propagation
information in its condition representation, which is consistent with the
performance gains observed in Table~\ref{tab:quantitative_comparison}.

\begin{table}[t]
\centering
\caption{Real-World Prediction Performance After Simulation-to-Real Adaptation.}
\label{tab:sim2real_performance}
\small  % 换成 small 字号，比 footnotesize 更清晰，比 normalsize 更紧凑
\setlength{\tabcolsep}{3pt} % 列间距微调到 5pt，保持左右平衡
\renewcommand{\arraystretch}{1.35} % 行高拉伸到 1.35，让表格看起来非常舒展
\begin{tabular}{lcccc}
\toprule
\textbf{Method}
& \textbf{Real Data}
& \textbf{NMSE} $\downarrow$
& \textbf{RMSE} $\downarrow$
& \textbf{MAE} $\downarrow$ \\
\midrule
Simulation Only
& No & 0.048 & 0.142 & 0.112 \\
LoRA~\cite{hu2021lora}
& Sparse & 0.036 & 0.128 & 0.097 \\
MAML~\cite{muruganandham2025smart}
& Sparse & 0.033 & 0.119 & 0.093 \\
Adversarial Learning~\cite{lu2026transfer}
& Sparse & 0.031 & 0.104 & 0.081 \\
\midrule
\textbf{DyTransRM (Ours)}
& \textbf{Sparse} & \textbf{0.023} & \textbf{0.079} & \textbf{0.064} \\
\bottomrule
\end{tabular}
\end{table}

We further evaluate whether the propagation semantics learned from simulation
remain effective under a simulation-to-real domain shift. For this zero-shot
experiment, all methods are trained exclusively on the RadioMapMotion
simulation dataset and directly evaluated on sparse real-world measurements,
without using any real-domain samples for parameter adaptation. The simulated
inputs are sparsified according to the same observation-density distribution
used for real-domain evaluation, and an identical measurement budget is
adopted for all methods. As shown in Table~\ref{tab:zero_shot_transfer}, all methods experience
performance degradation after direct transfer to the real domain, reflecting
the inevitable discrepancy between simulated and measured propagation
environments. Nevertheless, RadioCLIP retains the best zero-shot performance,
achieving an NMSE of 0.048, an RMSE of 0.142, and an MAE of 0.112.
Compared with the strongest baseline, RadioLSTM~\cite{cheng2026radiomapmotion},
RadioCLIP reduces the three errors by 9.4\%, 7.8\%, and 7.4\%, respectively.
The smaller performance margin than that in the simulation domain is expected
because real measurements contain propagation effects and environmental
discrepancies that are not fully reproduced by the simulator. More
importantly, the retained advantage indicates that the learned
physics-semantic representation is less dependent on simulation-specific RM
patterns, providing a transferable initialization for subsequent sparse
real-domain adaptation.

\subsection{Simulation-to-Real Adaptation Performance}
\label{subsec:sim2real_performance}

We next evaluate the complete DyTransRM framework under sparse real-domain
supervision. Unlike the zero-shot setting, a limited number of measurements
from the target-domain training split are available for adaptation, while the
test set remains strictly held out. All adaptation methods start from the same
simulation-trained predictor and use the same target-domain measurements and
evaluation protocol, such that the comparison isolates the effectiveness of
different simulation-to-real adaptation strategies.

We first evaluate the performance of the proposed framework on the Vienna dataset. As shown in Table~\ref{tab:sim2real_performance}, sparse target-domain
adaptation substantially improves prediction accuracy over direct
simulation-to-real transfer. DyTransRM achieves the best performance, with
NMSE, RMSE, and MAE of 0.023, 0.079, and 0.064, respectively. Compared with
the strongest competing adaptation method, adversarial learning, DyTransRM
reduces the three errors by 25.8\%, 24.0\%, and 21.0\%, respectively.
Relative to the simulation-only model, the reductions reach 52.1\%, 44.4\%,
and 42.9\%. The improvement stems from the complementary roles of sparse-mask refinement
and UOT-based alignment. Sparse-mask refinement exposes the predictor to
randomly varying receiver configurations during simulation training, reducing
the mismatch between densely available simulated RMs and sparsely observed
real measurements. UOT then operates in the learned propagation-semantic
space to align simulation and real representations while relaxing the
requirement that all samples contribute equally to cross-domain matching.
Consequently, simulation samples that are compatible with the target-domain
propagation patterns contribute more strongly to knowledge transfer, whereas
poorly matched simulation-specific patterns exert less influence. These results show that DyTransRM can effectively exploit sparse
real measurements to compensate for the simulation-to-real discrepancy
without requiring spatially complete real RMs.

\begin{figure}[t]
\centering  
\includegraphics[height=4.8cm,width=8.8cm]{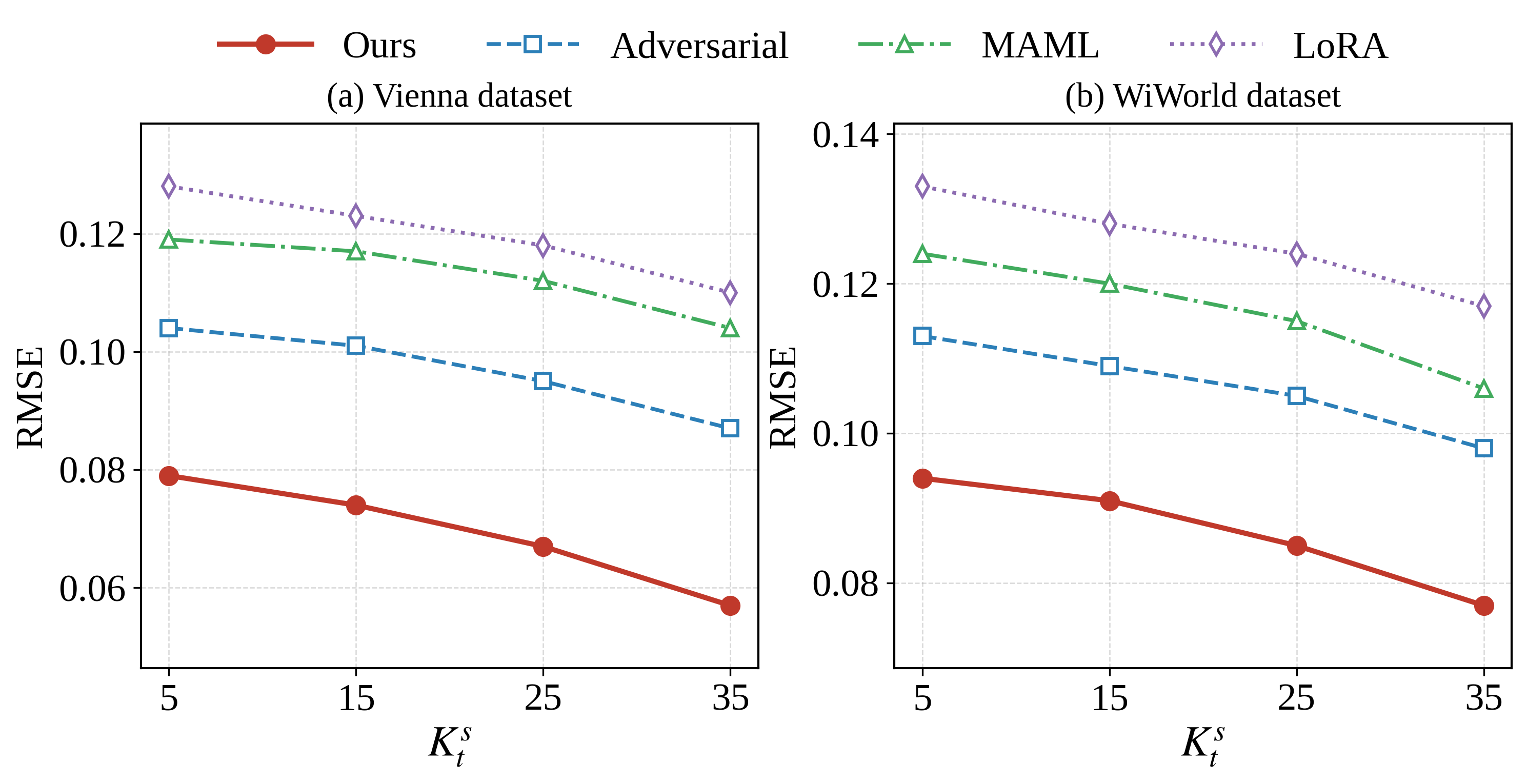}
\caption{RMSE performance under different numbers of sampled points per time slot on the (a) Vienna and (b) WiWorld datasets.}
\label{fig:training_compare}
\end{figure}

Fig.~\ref{fig:training_compare} further evaluates robustness to measurement
sparsity by varying the number of observed receiver locations per time slot.
On both the Vienna and WiWorld datasets, the RMSE decreases as more
measurements become available, as expected. More importantly, DyTransRM
consistently achieves the lowest RMSE across all considered sampling budgets.
Its advantage is already evident with only five observed locations per time
slot and persists as the number of measurements increases to 35. This
consistent behavior indicates that the proposed adaptation strategy does not
depend on a particular observation density and remains effective under
severely sparse and time-varying receiver measurements.

\subsection{Ablation and Efficiency Analysis}
\label{subsec:ablation}

We first conduct an ablation study to evaluate the contribution of each component in DyTransRM. 
Starting from the same spatio-temporal backbone, RadioCLIP conditioning, sparse-mask refinement, and UOT-based domain alignment are progressively introduced, while the remaining settings are kept unchanged.

As shown in Table~\ref{tab:ablation}, the backbone achieves an NMSE, RMSE, and MAE of 0.055, 0.156, and 0.121, respectively. 
Introducing RadioCLIP reduces them to 0.048, 0.142, and 0.112, demonstrating the benefit of incorporating environment-aware propagation semantics into dynamic RM prediction. 
Sparse-mask refinement further improves the results to 0.036, 0.113, and 0.089, indicating that explicitly simulating sparse observations effectively reduces the mismatch between dense simulated RMs and sparse real-world measurements. 
Finally, UOT alignment achieves the best performance of 0.023, 0.079, and 0.064, corresponding to further reductions of 36.1\%, 30.1\%, and 28.1\% over the preceding configuration. 
These results show that sparse-mask refinement and UOT address complementary aspects of the simulation-to-real gap: the former alleviates observation-pattern mismatch, whereas the latter aligns cross-domain propagation representations.

We further evaluate the computational efficiency of DyTransRM in Table~\ref{tab:efficiency}. 
DyTransRM contains 17.39~M parameters and requires 243.39~G FLOPs, with an average inference latency of 46.81~ms. 
Compared with RadioLSTM, it reduces FLOPs by 75.4\% while maintaining comparable latency, and compared with BERT, it uses 61.5\% fewer parameters and reduces latency from 58.53~ms to 46.81~ms. 
Although CSI-MAE and JEPA-CFM are computationally lighter, their prediction accuracy is considerably lower, revealing a clear accuracy--efficiency tradeoff.

Importantly, RadioCLIP pretraining and UOT-based adaptation are performed offline and introduce no iterative optimization during inference. 
Therefore, the proposed framework improves prediction and simulation-to-real adaptation without substantially increasing online inference overhead, supporting its practical use for dynamic RM forecasting.

\begin{table}[t]
\centering
\caption{Ablation Study of the Proposed DyTransRM Framework}

\label{tab:ablation}
\footnotesize
\setlength{\tabcolsep}{3.3pt}
\renewcommand{\arraystretch}{1.35} % 行高拉伸到 1.35，让表格看起来非常舒展
\begin{tabular}{ccccccc}
\toprule
\textbf{Backbone}
& \textbf{RadioCLIP}
& \textbf{Mask}
& \textbf{UOT}
& \textbf{NMSE} $\downarrow$
& \textbf{RMSE} $\downarrow$
& \textbf{MAE} $\downarrow$ \\
\midrule
\checkmark &           &           &           & 0.055 & 0.156 & 0.121 \\
\checkmark & \checkmark&           &           & 0.048 & 0.142 & 0.112 \\
\checkmark & \checkmark& \checkmark&           & 0.036 & 0.113 & 0.089 \\
\checkmark & \checkmark& \checkmark& \checkmark& \textbf{0.023} & \textbf{0.079} & \textbf{0.064} \\
\bottomrule
\end{tabular}
\end{table}

\begin{table}[!t]
  \centering
  \caption{Computational Complexity and Inference Efficiency.}
  \label{tab:efficiency}
  \renewcommand{\arraystretch}{1.35} % 行高拉伸到 1.35，让表格看起来非常舒展
  \begin{tabular}{lccc}
    \toprule
    Method & Params. (M) & FLOPs (G) & Latency (ms)~$\downarrow$ \\
    \midrule
    RadioLSTM & 2.69 & 990.90 & 48.79 \\
    BERT & 45.15 & 224.99 & 58.53 \\
    CSI-MAE & 3.97 & 7.57 & 2.95 \\
    JEPA-CFM & 8.84 & 15.99 & 8.13 \\
    \midrule
    DyTransRM (Ours) & 17.39 & 243.39 & 46.81 \\
    \bottomrule
  \end{tabular}
\end{table}

\section{CONCLUSION}
\label{sect:conc}
In this paper, we proposed DyTransRM, a unified simulation-to-real framework for dynamic RM prediction under sparse real-world measurements. To improve the transferability of simulation-learned propagation knowledge, RadioCLIP explicitly associates environmental evolution with propagation evolution through cross-modal contrastive learning and incorporates the resulting physics-semantic representation into spatio-temporal RM prediction. To address the remaining simulation-to-real discrepancy, SparseUOT combines sparse-mask simulation refinement with unbalanced optimal transport, jointly alleviating the observation-pattern mismatch and propagation-distribution mismatch between the two domains. Experimental results demonstrate that RadioCLIP improves dynamic RM prediction and retains stronger zero-shot transferability than existing baselines. With sparse real-domain adaptation, DyTransRM further achieves an NMSE of 0.023, RMSE of 0.079, and MAE of 0.064, while requiring no spatially complete real RM supervision. The consistent improvements under different measurement densities further validate its robustness to sparse observations. Moreover, the proposed cross-modal pretraining and domain alignment are performed offline, preserving practical online inference efficiency. These results demonstrate that combining physics-semantic learning with sparse observation-aware domain adaptation provides an effective approach for transferring large-scale simulated propagation knowledge to real-world dynamic RM prediction.

\bibliographystyle{IEEEtran}
\bibliography{references}
\end{document}